\documentclass[prb, preprint, amsmath, amssymb, showpacs, superscriptaddress]{revtex4-1}
\usepackage[T1]{fontenc}
\usepackage{lmodern}
\usepackage{graphicx} 
\usepackage{dcolumn}  
\usepackage{bm}       
\usepackage{amsfonts}
\usepackage{dsfont}
\usepackage{csquotes}
\usepackage{tikz}
\usepackage{ulem}
\usepackage{color}

\usepackage{physics}
\usepackage{hyperref}

\begin{document}

\title{Helicity-engineered nonlinear optical responses in photo-excited topological semimetals}

\author{Roshan Kumar Thakur}
 \affiliation{Department of Physics, Indian Institute of Technology Bombay, Powai, Mumbai-400076, India}
 
\author{Prachi Venkat}%
\affiliation{Department of Physics, Indian Institute of Technology Bombay, Powai, Mumbai-400076, India}%

\author{Amar Bharti}
\affiliation{Collective Dynamics and Quantum Transport Unit,
OIST Graduate University, Onna 904 0495, Japan}%

\author{Gopal Dixit}
\affiliation{Department of Physics, Indian Institute of Technology Bombay, Powai, Mumbai-400076, India}%

\date{\today}

\begin{abstract}

Topological materials provide a transformative arena for light-driven control of electronic motion; yet, direct manipulation of electron dynamics on sub-cycle timescale remains a significant challenge. 
We demonstrate that the strong-field-driven high-harmonic generation  of a Weyl semimetal  
can be controlled and manipulated through bicircular pump-probe driving fields. 
While a lone probe pulse generates exclusively  odd-order harmonics, the addition of a pump pulse triggers a series of sidebands arising from nonlinear frequency mixing of pump and probe photons. 
Our results reveal that the sideband intensities are highly sensitive to the relative helicity of the pulses and the orientation of the polarization plane. 
This sensitivity stems from the chiral nature of the Weyl nodes, which couples efficiently to the light's helicity only when the node-separation axis is perpendicular to the polarization plane.
Furthermore, the significant suppression of sideband intensity with increasing pump-probe delay identifies these features as a potential  clock for electron-hole decoherence. 
These findings establish frequency-mixed high-harmonic generation as a sensitive  probe of chiral quantum dynamics and suggest a robust framework for manipulating topological currents via structured light, with implications for lightwave electronics and ultrafast quantum information processing.

\end{abstract}

\maketitle

\section{Introduction}

The emergence of topological materials has fundamentally reshaped our understanding of contemporary physics while providing a robust platform for next-generation quantum technologies~\cite{hasan2010colloquium, keimer2017physics, xu2015discovery1}. 
These materials are characterized by nontrivial band topologies and symmetry-protected states that manifest in exotic electronic phenomena. 
Critically, their light-matter interactions are dictated by the geometric properties of the electronic wavefunction~\cite{Xiao2010}, enabling unconventional transport and nonlinear optical responses under coherent excitation~\cite{ bao2021light, vazifeh2013electromagnetic,  meng2019large, yang2015chirality, burkov2014chiral, trescher2015quantum}.
Within this landscape, topological semimetals  have emerged as a pivotal class due to their highly tunable electronic structures and complex symmetry manifolds. 
Of particular interest are Weyl semimetals (WSMs), where the lifting of either inversion or time-reversal symmetry permits the existence of Weyl nodes. 
These Weyl nodes, occurring at the linear crossings of nondegenerate bands, act as quantized 
sources and sinks -- monopoles -- of Berry curvature in momentum space~\cite{Armitage2018, yan2017topological}.

Parallel to the maturation of topological materials, 
the development of tunable, intense ultrafast laser sources has unlocked the ability to interrogate solids across the perturbative and nonperturbative regimes. 
By precisely tailoring the waveform and intensity of these ultrashort pulses, it is now possible to drive and manipulate electron dynamics in previously inaccessible  regimes of light-matter interaction. 
A seminal milestone in this field was the observation of nonperturbative optical responses in ZnO~\cite{ghimire2011observation}, which established high-harmonic generation (HHG) as a versatile spectroscopic tool for probing nonequilibrium phenomena across diverse condensed matter systems~\cite{ghimire2019, vampa2015all, hohenleutner2015real, tancogne2017ellipticity, liu2018enhanced,  gauthier2019orbital, imai2020high, yue2020imperfect,  borsch2020super,  shcherbakov2021generation,  rana2022probing, kaassamani2022polarization, qian2022role, rana2022generation, cha2022gate, shi2023giant, li2023high, avetissian2024berry, rana2024high, kim2024dephasing}.
Recently, research has converged at the intersection of HHG and topological matter, with a burgeoning focus on WSMs~\cite{chacon2020circular, cheng2020efficient,  lim2020efficient, lv2021high, schmid2021tunable, baykusheva2021all,  heide2022probing,  li2022high, avetissian2022high, bharti2022high,   bharti2023role,  neufeld2023there, graml2023influence, wang2024high, wang2024table, bharti2024non, medic2024high, liu2025theoretical, yao2025high, lorenzo2025high, venkat2026attosecond, li2026disentangling}. 
Despite this interest, the fundamental nature of the nonperturbative response in photoexcited WSMs remains largely unmapped. 
HHG in these systems offers a unique spectroscopic window into ultrafast electronic dynamics dictated by nontrivial band geometry. Specifically, pump-probe architectures provide a robust framework to access these non-equilibrium states, wherein a primary pump pulse initializes a carrier distribution that is subsequently interrogated by a strong-field probe.
This approach enables the real-time observation and control of topological charge dynamics on subcycle timescale.

\begin{figure}[h!]
 \centering
 \includegraphics[width= \linewidth]{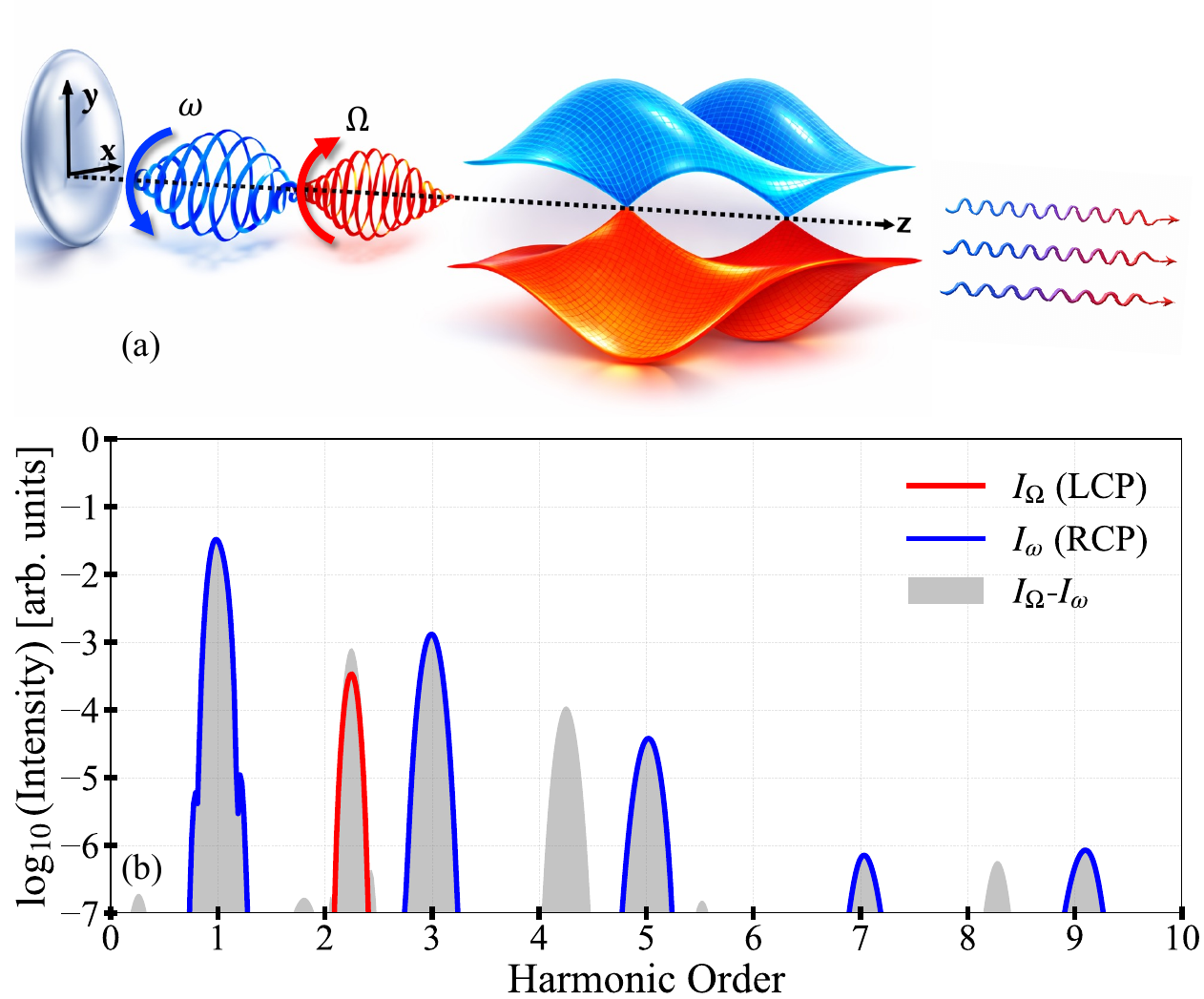} 
 \caption{(a) Schematic illustration of the interaction of two-color circularly polarized laser pulses with an inversion-symmetric Weyl semimetal in a pump-probe configuration.
(b) High-harmonic spectra generated by the pump pulse ($\Omega$, in red) and probe pulse 
($\omega$, in blue), along with the spectrum obtained from the combined pump-probe excitation (gray). The pump pulse is left-circularly polarized (LCP), while the probe pulse is right-circularly polarized (RCP). 
The pump and probe wavelengths are chosen to be incommensurate in order to avoid spectral overlap of their respective responses.
The harmonic orders are expressed in units of the probe frequency $\omega$.}
  \label{fig1}
\end{figure}

The present work investigates the role of photo-doping on HHG in a WSM. 
To this end, a pump pulse is employed to pre-excite carriers near the chiral Weyl nodes, 
followed by a strong probe pulse that drives non-perturbative harmonic emission, as schematically illustrated in Fig.~\ref{fig1}(a). 
The pump ($\Omega$) and probe ($\omega$) frequencies are chosen to be incommensurate to prevent spectral overlap between their respective contributions.
We systematically examine all helicity combinations for the circularly polarized pump and probe fields. 
Our results demonstrate that the photoexcited WSM exhibits pronounced nonlinear frequency mixing between the pump and probe photons, resulting in the emergence of distinct sidebands in the harmonic spectrum. 
Furthermore, this frequency mixing is highly sensitive to both the relative polarization and the temporal delay between the pulses. 
This sensitivity provides a robust spectroscopic means to extract critical information regarding electron-hole coherence, specifically the dephasing time of the system.

\section{Methodology}
The electronic structure of an inversion-symmetric WSM 
is described within a tight-binding framework by the Hamiltonian as
$\mathcal{H} =  \mathbf{d}(\mathbf{k}) \cdot \sigma$,  
where $\sigma$ denotes the Pauli matrices. 
Following the model in Ref.~\cite{bharti2023tailoring}, the components of the vector $\mathbf{d}(\mathbf{k})$
are defined as
$\mathbf{d}(\mathbf{k}) = [t\sin(k_{x}a), t\sin(k_{y}a), t\{\cos(k_{z}a) - \cos(k_{0}a) + 2 - \cos(k_{x}a) - \cos(k_{y}a) \}]$. 
This inversion-symmetric WSM hosts two Weyl nodes located at $\mathbf{k} = [0, 0, \pm\pi/(2a)]$ with 
$a = 6.28$ \AA~ as the lattice constant of the cubic lattice. 
Throughout this work, we consider an isotropic hopping parameter $ t = $ 1.8 eV.
A weak pump pulse with a wavelength of 800 nm (1.55 eV) and peak intensity of $5\times10^{9}$ W/cm$^2$ 
 is used to photoexcite the WSM, while an intense probe pulse with a wavelength of 1800 nm 
 (0.69 eV) and peak intensity of $5\times10^{11}$ W/cm$^2$
drives the nonlinear optical response from the photoexcited WSM. 
Both pulses have a duration of approximately 100 fs and are temporally overlapped, unless stated otherwise. 

The intense-laser-driven electronic dynamics of the WSM in a pump-probe configuration are modeled by solving the Heisenberg equation of motion within the density-matrix formalism in the Houston basis~\cite{wilhelm2021semiconductor, yue2022introduction}. 
The equation is  solved numerically in the velocity gauge using a fourth-order Runge-Kutta scheme 
with a uniform $90 \times 90 \times 90~\textbf{k}$-point grid to sample the Brillouin zone and a temporal step of $ \Delta t = 14$ as, following the approach described in Refs.~\cite{mrudul2021high, rana2022high}.
The high-harmonic spectrum is calculated from the Fourier transform of the time-derivative of the electronic current, given by $|\mathcal{FT}(\partial \mathbf{J}(t)/\partial t)|^2$, where $\mathbf{J}(t)$  
denotes the total electronic current induced by the driving field.
Decoherence of electron and hole 
is incorporated through a phenomenological dephasing time $\textrm{T}_2$  = 1.5 fs ~\cite{bharti2024non}.

\section{Results and discussion}

Owing to its relatively low intensity, the pump pulse alone does not generate significant higher-order nonlinear responses, as shown in Fig.~\ref{fig1}(b). 
In contrast, the probe pulse is sufficiently intense to produce higher-order harmonics 
up to the 9$^{\textrm{th}}$ order [blue curve in Fig.~\ref{fig1}(b)]. 
Furthermore, the generated harmonics are exclusively odd-order, consistent with the inversion symmetry of the WSM. 
The spectrum in Fig.~\ref{fig1}(b) changes markedly when the pump and probe pulses interact simultaneously. 
In this case, additional spectral peaks emerge [gray curve in Fig.~\ref{fig1}(b)], beyond those produced by the individual pump and probe pulses. 
The energetic positions of these additional peaks indicate the generation of sum- and difference-frequency components arising from the nonlinear mixing of pump and probe photons.
Consequently, the interaction of pump and probe pulses with opposite helicities results in efficient frequency mixing and the emergence of distinct sidebands in the high-harmonic spectrum. 
Motivated by these observations, we now investigate how the frequency mixing is affected by different combinations of pump-probe helicities.

\begin{figure}[h!]
\includegraphics[width=\textwidth]{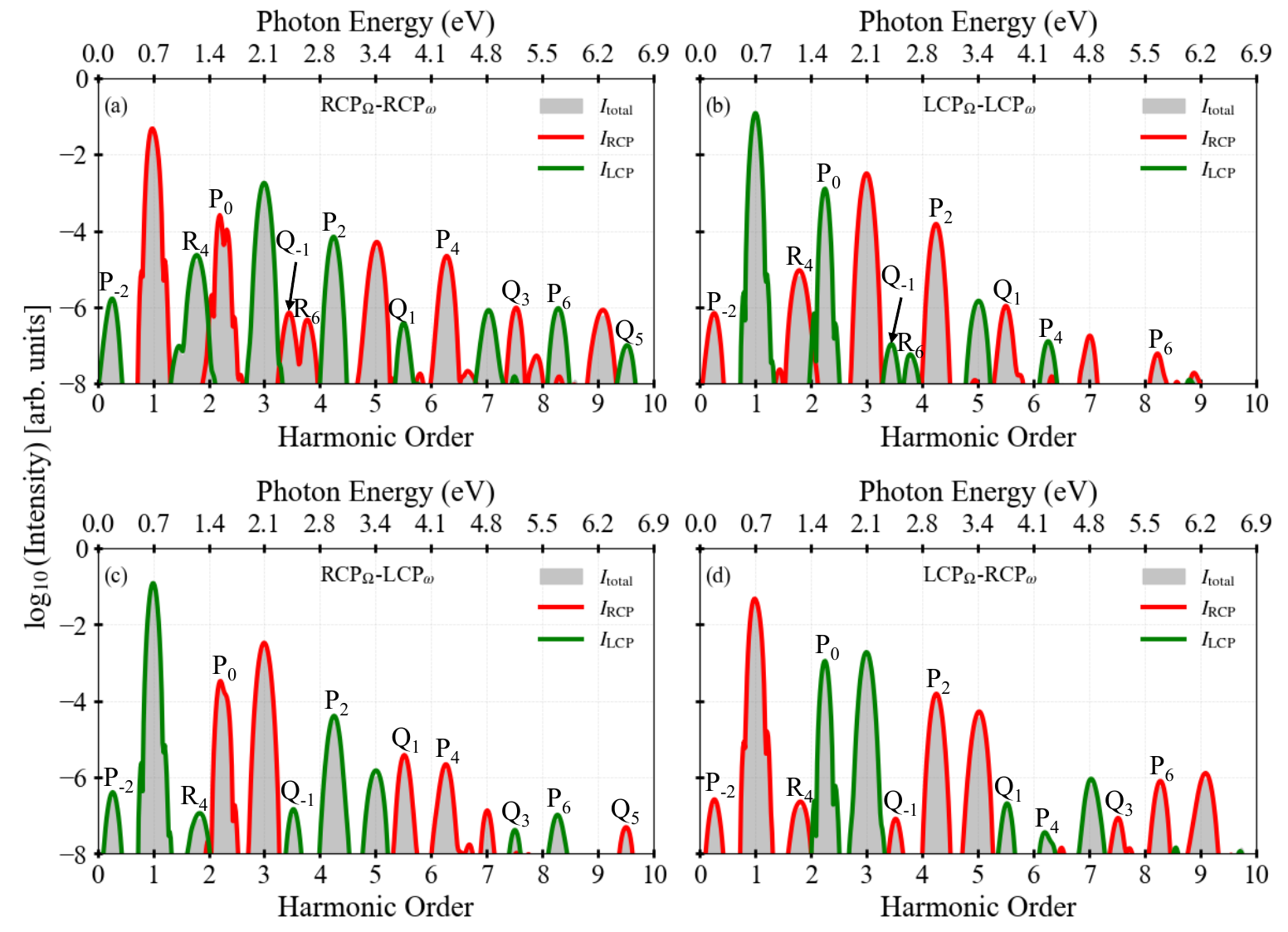}
\caption{High-harmonic spectra for different polarization configurations of the pump ($\Omega$) 
and probe ($\omega$) pulses: (a) RCP$_\Omega$-RCP$_\omega$, 
(b) LCP$_\Omega$-LCP$_\omega$,  (c) RCP$_\Omega$-LCP$_\omega$, and (d) LCP$_\Omega$-RCP$_\omega$. 
The total harmonic and sideband spectra are shown by the gray shaded region, while the helicity-resolved contributions corresponding to right- and left-circular polarization (RCP and LCP) 
are indicated by the red and green curves, respectively. 
The pump and probe photon energies are 1.55 eV (800 nm) and 0.69 eV (1800 nm), respectively. 
The harmonic orders are reported in units of the probe frequency.}
 \label{fig2}
\end{figure}

By comparing the spectral positions of the sidebands with the primary harmonic peaks, 
clear signatures of frequency mixing are observed for all pump-probe helicity combinations, 
as illustrated in Fig.~\ref{fig2}. 
Depending on the number of pump and probe photons involved, three classes of mixing processes can be identified, denoted as $\mathcal{P, Q}$ and $\mathcal{R}$. 
Specifically, class $\mathcal{P}$ is defined by $\mathcal{P}_p = \{p\omega + \Omega~|~p \in \{-2, 0, 2, 4, 6\}\}$, where $\mathcal{P}_0$ represents the fundamental frequency of the pump pulse; 
class $\mathcal{Q}$ follows $\mathcal{Q}_q=\{q\omega + 2\Omega~|~q \in \{-3,-1, 1, 3\}\}$; 
and class $\mathcal{R}$ is given by $\mathcal{R}_r=\{r\omega - \Omega ~|~ r \in \{4, 6\}\}$. 
Notably, in all three cases, the sidebands emerge from nonlinear mixing processes involving an odd total number of $\Omega$  and $\omega$  photons. 
Furthermore, both the intensities and polarization states of the generated primary harmonic peaks and sidebands exhibit a sensitive dependence on the relative helicities of the pump and probe fields.

For instance, the $9^{\textrm{th}}$ harmonic is observed when the $\omega$ pulse is RCP [see Figs.~\ref{fig2}(a) and \ref{fig2}(d)], but becomes suppressed when the polarization changes from RCP to LCP [see Figs.~\ref{fig2}(b) and \ref{fig2}(c)].
Sidebands up to $\mathcal{Q}_5$ are clearly visible for the RCP-RCP and RCP-LCP combinations [Figs.~\ref{fig2}(a) and \ref{fig2}(c)], while the sideband  $\mathcal{P}_6$  appears in all polarization configurations. 
However, the intensity of $\mathcal{P}_6$  varies by nearly an order of magnitude depending on the helicity combination of the driving fields. Notably, $\mathcal{P}_6$  
exhibits an intensity comparable to that of the $9^{\textrm{th}}$ harmonic in Figs.~\ref{fig2}(a) and \ref{fig2}(d), whereas the $\mathcal{P}_6$ sideband remains allowed even though the $9^{\textrm{th}}$  harmonic is suppressed  in Figs.~\ref{fig2}(b) and \ref{fig2}(c).
Similarly, the sideband $\mathcal{Q}_3$  has an intensity comparable to that of the $7^{\textrm{th}}$ harmonic in Figs.~\ref{fig2}(a) and \ref{fig2}(c), 
but is absent or significantly weaker in Figs.~\ref{fig2}(b) and \ref{fig2}(d), respectively.
A comparison of Figs.~\ref{fig2}(a) and \ref{fig2}(d) further reveals that changing the pump's helicity  
leads to substantial variations in the sideband intensities. 
This behavior is characteristic of WSM with chiral Weyl nodes and can be attributed to circular dichroism, whereby pump pulses of opposite helicities selectively photoexcite different regions of the electronic band structure in the vicinity of the Weyl nodes.

We now turn to the polarization properties of the emitted spectra. 
Owing to the fourfold rotational symmetry of the cubic lattice of the WSM, the harmonics generated by the 
$\omega$-field obey the  ($4m \pm 1$)-selection rule with  $m$ as integer, 
resulting in an alternating helicity pattern in the emitted spectrum, as shown in Fig.~\ref{fig2}.
The fundamental harmonic in each case inherits the helicity of the $\omega$ pulse, while the subsequent harmonics alternate in helicity with increasing order. 
In contrast, the sidebands arising from frequency mixing involve processes with an odd total number of photons and therefore depend sensitively on the polarization states of both the $\Omega$ and $\omega$ pulses.

The sidebands arise from nonlinear frequency-mixing processes involving the absorption of multiple photons from the $\Omega$- and $\omega$-fields, followed by emission of a harmonic photon. 
Each circularly polarized photon carries a spin angular momentum  of $\pm 1$, 
corresponding to RCP and LCP. 
Angular momentum conservation in the sideband generation process can therefore be written as
\begin{equation}\label{eq:rule}
n_{\Omega}\sigma_{\Omega} + n_{\omega}\sigma_{\omega} = 4m \pm 1, 
\end{equation}
where $n_{\Omega}$ and $n_{\omega}$ denote the number of absorbed $\Omega$ and $\omega$ photons, respectively.
The term $4m$ reflects the exchange of an angular momentum in multiples of $4 \hbar$, 
imposed by the fourfold rotational symmetry of the cubic lattice.
Here, $\sigma_{\Omega}, \sigma_{\omega} = \pm 1$ denote the helicities of the driving fields.

To satisfy the requirement that an odd total number of photons participates in the sideband generation process, consider the case where both driving fields are RCP, 
i.e., $\sigma_{\Omega} = \sigma_{\omega} = +1$.
In this case, absorption of two  $\omega$-photons and one $\Omega$-photon results in a total angular momentum of +3, corresponding to the sideband $\mathcal{P}_2 = 2\omega + \Omega$. 
If the emitted photon has LCP (-1), 
the angular momentum conservation condition yields $4m - 1 = +3$ for $m = 1$, demonstrating that this mixing process satisfies the ($4m \pm 1$)-selection rule and is therefore symmetry-allowed.
According to Eq.~(\ref{eq:rule}), consecutive sidebands in each of the  $\mathcal{P, Q}$, and $\mathcal{R}$ series possess alternating helicities. 
Furthermore, the helicity of the sidebands in the $\mathcal{P}$ series reverses when the helicity of the $\Omega$ field changes from RCP to LCP, as shown in Figs.~\ref{fig2}(a) and \ref{fig2}(d). 
In contrast, frequency mixing in inversion-symmetry-broken materials, such as monolayer MoS$_2$, does not display such helicity-dependent behavior~\cite{venkat2026}.

To illustrate how angular momentum conservation forbids certain sideband generation processes, let us consider the frequency mixing involving the absorption of one photon from each of the 
$\Omega$- and $\omega$-fields with both RCP, i.e., $\sigma_{\Omega} = \sigma_{\omega} = +1$.
According to Eq.~(\ref{eq:rule}), the total angular momentum of the absorbed photons is 1(+1)+1(+1)=+2. 
However, the selection rule requires that 
$n_{\Omega} \sigma_{\Omega} + n_{\omega}\sigma_{\omega} = 4m \pm 1$, 
which implies that the allowed angular momenta must follow the series 
$4m + 1 = 1, 5, 9, \dots$ or $4m - 1 = 3, 7, 11, \dots$
Since the value +2 does not satisfy the condition $4m \pm 1$  for any integer $m$, 
this frequency-mixing process is symmetry-forbidden. 
Consequently, the corresponding sideband is absent in the spectrum, consistent with the results shown in Fig.~\ref{fig2}.

\begin{figure}[h!]
\includegraphics[width=\textwidth]{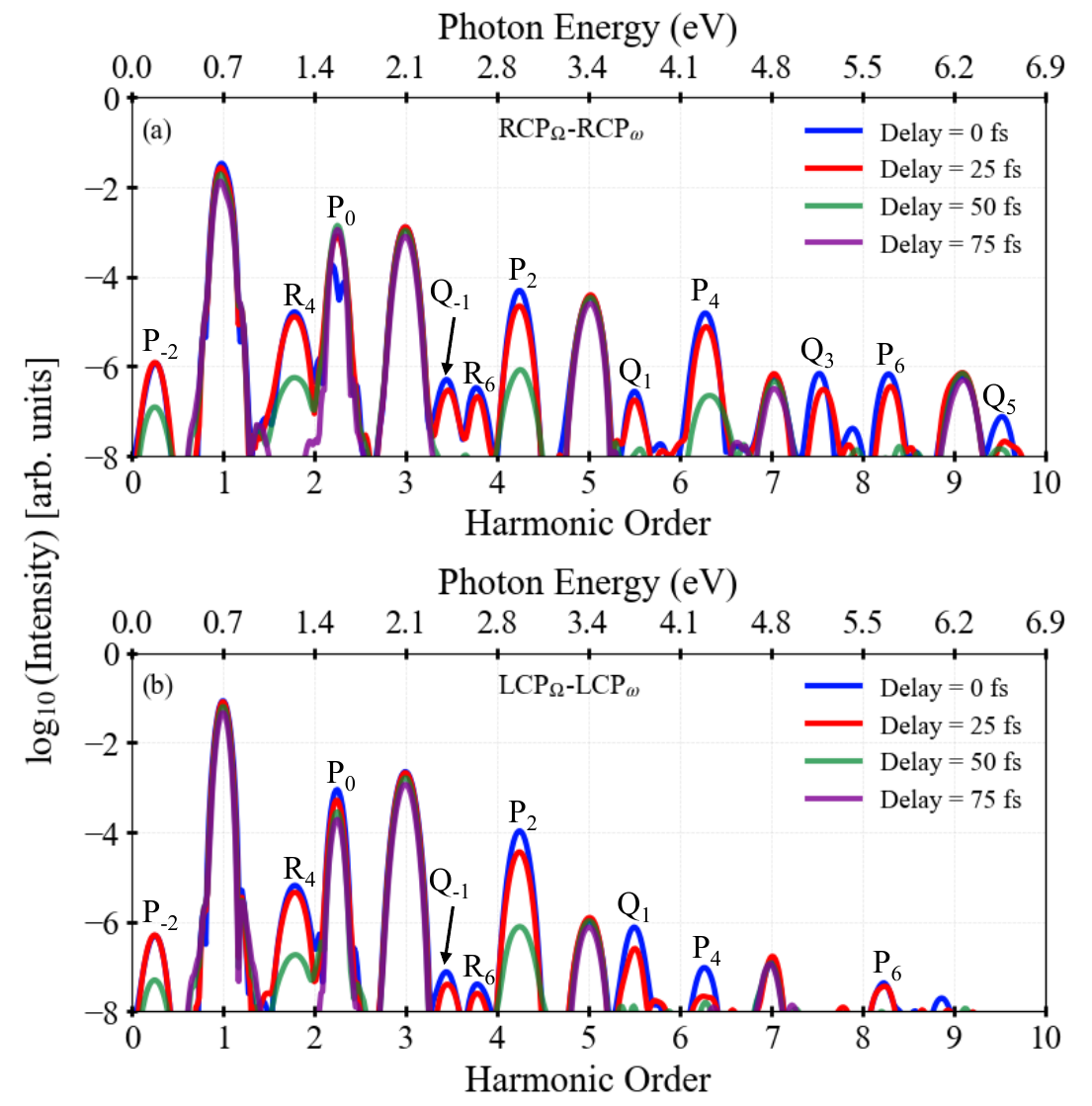}
\caption{Dependence of the high-harmonic response on the relative delay between the pump and probe pulses for the polarization combinations (a) RCP$_\Omega$-RCP$_\omega$ and (b) LCP$_\Omega$-RCP$_\omega$.}
\label{fig3}
\end{figure}

So far, our discussion has focused on the case where the pump and probe pulses temporally overlap. 
It is imperative to examine how the relative delay between the pump and probe pulses influences the nonlinear optical response of the photoexcited WSM. 
To illustrate the role of the delay, we consider two representative polarization configurations. 
the sideband intensities remain largely unaffected when the delay between the pump and probe pulses is increased to 50 fs ($\sim 8$ optical cycles of $\Omega$ pulse) as shown in Figs.~\ref{fig3}(a) and \ref{fig3}(b) for the RCP$_\Omega$-RCP$_\omega$ and LCP$_\Omega$-LCP$_\omega$ combinations, respectively. 
However, a further increase in the delay to 75 fs ($\sim 12$ cycles of $\Omega$ pulse) 
leads to a pronounced suppression of the sideband intensity. 
In contrast, the main harmonic peaks remain largely insensitive to the  delay. 
The relative robustness of the main harmonic signal against delays of several hundreds of femtoseconds has been reported previously in both experimental and theoretical studies of solids~\cite{venkat2026, Heide2022, Peterka2023}. 
The strong sensitivity of the sideband intensity to pump-probe delays on the order of a few tens of femtoseconds suggests that these features can serve as a probe of the electron-hole coherence.

\begin{figure}[h!]
\includegraphics[width=\textwidth]{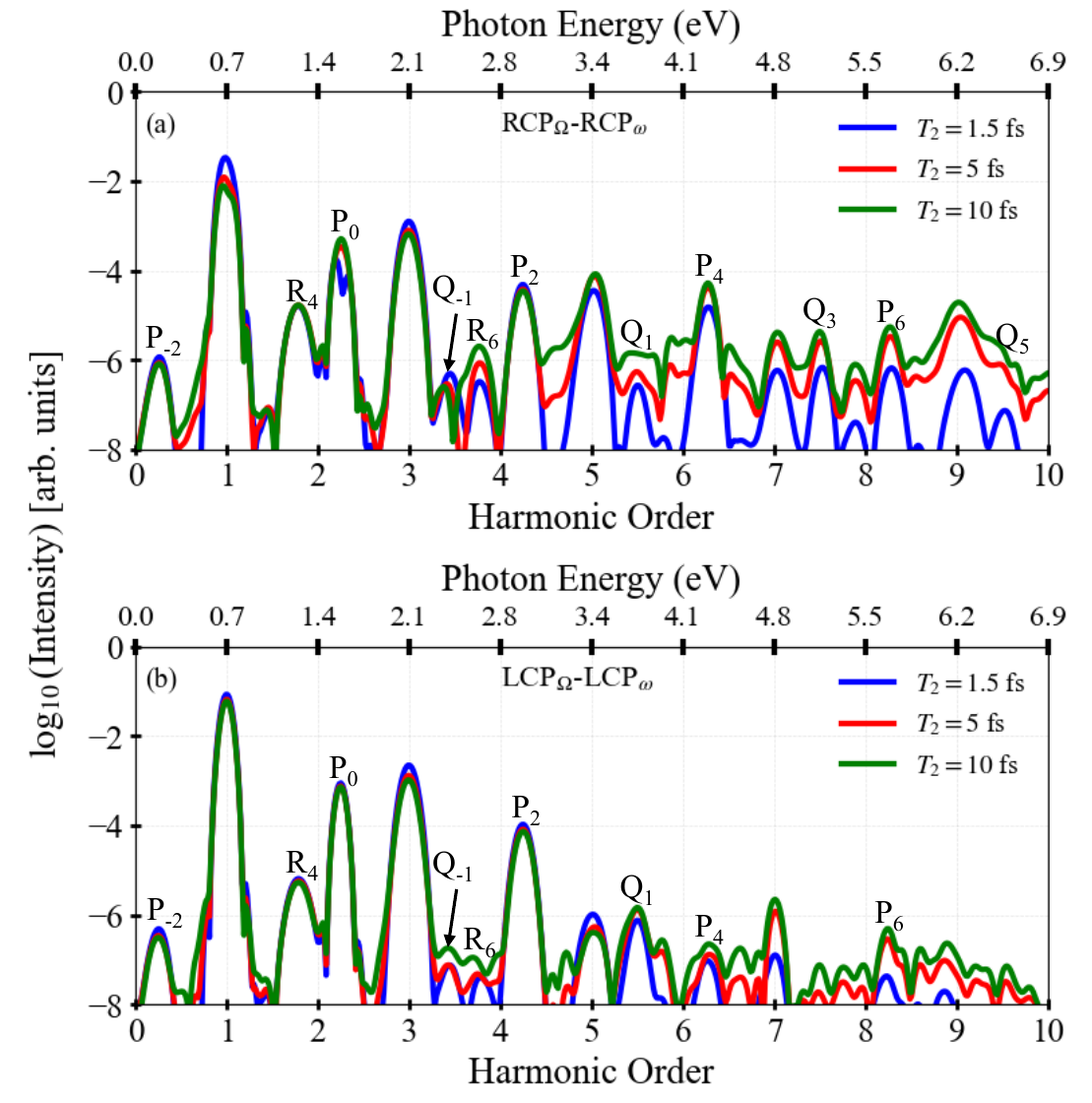}
\caption{Influence of the dephasing time ($\textrm{T}_2$) on the high-harmonic spectra for the polarization configurations (a) RCP$_\Omega$-RCP$_\omega$ and (b) LCP$_\Omega$-LCP$_\omega$, highlighting the role of electron-hole coherence.} \label{fig4}
\end{figure}

Decoherence between electrons and holes is incorporated phenomenologically through the dephasing time 
$\textrm{T}_2$ in the present work. 
It is therefore important to examine how the choice of dephasing time influences our findings. 
Figure~\ref{fig4} illustrates the $\textrm{T}_2$-dependence of the spectra for two representative polarization configurations -- RCP$_{\Omega}$-RCP$_{\omega}$  and LCP$_{\Omega}$-RCP$_{\omega}$ -- 
with $\textrm{T}_2$ values ranging from 1.5 to 10~fs. 
An increase in $\textrm{T}_2$ signifies prolonged electron-hole coherence, 
which facilitates an enhancement of the higher-order harmonics primarily driven by interband transitions.
A comparable trend is observed for the sidebands; specifically, the intensities of the $\mathcal{Q}_1, \mathcal{Q}_3, \mathcal{P}_4$ and $\mathcal{P}_6$  sidebands increase by at least an order of magnitude in both polarization configurations.
This enhancement can be attributed to the longer coherence time, which increases the probability of coherent electron-hole recombination along their laser-driven trajectories.

\begin{figure}[h!]
\includegraphics[width=\textwidth]{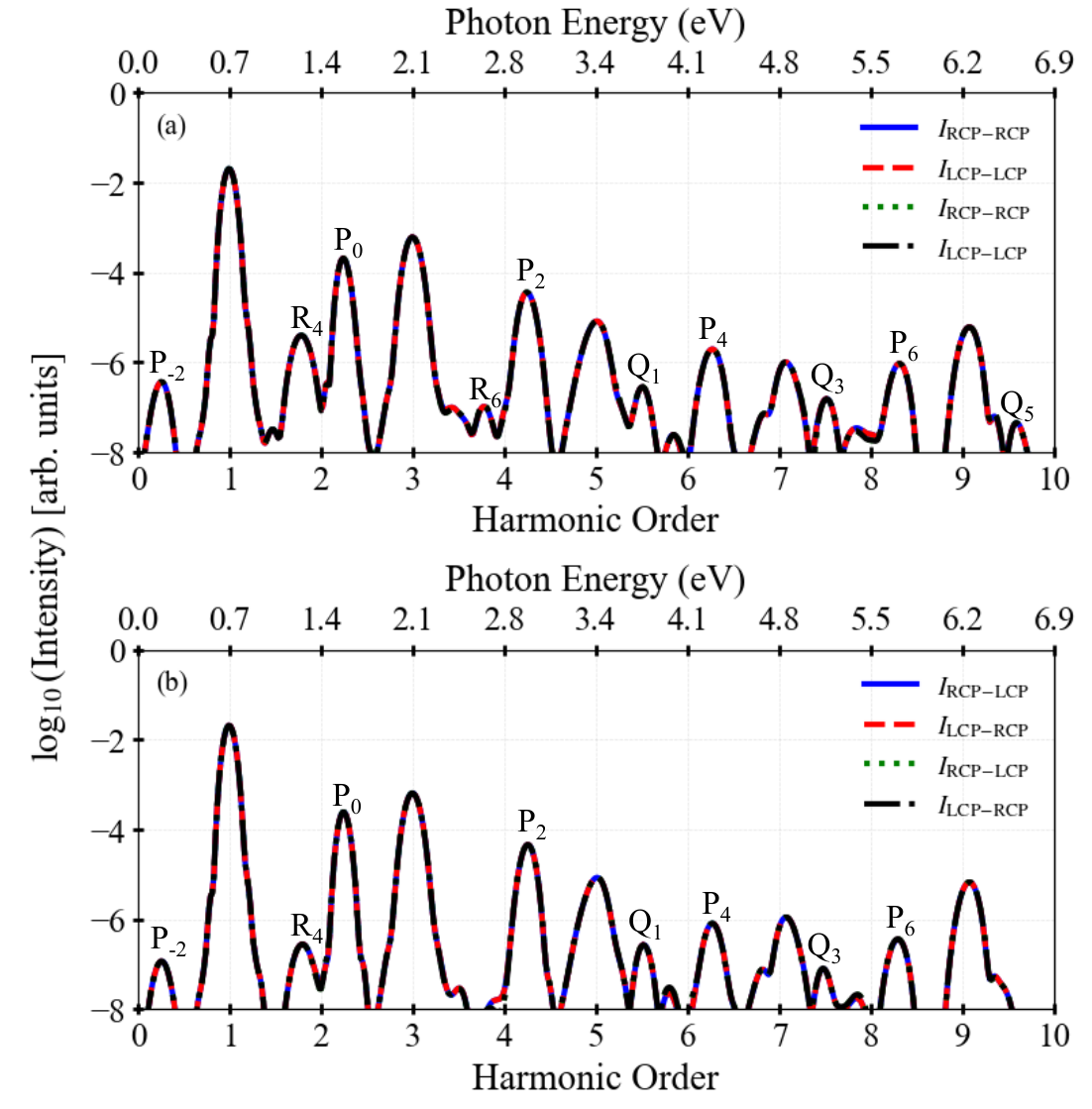}
 \caption{High-harmonic spectra for (a) RCP-RCP and LCP-LCP pump-probe polarization configurations resolved in the $x-z$ plane (blue and red curves) and the $y-z$ plane (green and black curves).
(b) Same as in (a), but for the RCP-LCP and LCP-RCP polarization combinations of the pump-probe pulses.} \label{fig5}
\end{figure}

Before concluding, we examine how the polarization plane of the pump-probe pulses influences the nonlinear frequency mixing. 
So far, our discussion has focused on the case where both circularly polarized pump and probe pulses lie in the $x-y$  plane. 
When the pump and probe pulses with identical helicities are polarized in either the $x-z$  or $y-z$  planes, the resulting spectra remain essentially identical as shown in Fig.~\ref{fig5}(a).
Similarly, opposite-helicity combinations of the pump and probe pulses in planes containing the $z$-axis also produce nearly identical spectra as illustrated in Fig.~\ref{fig5}(b).
In all cases, both the main harmonic peaks and the corresponding sidebands are  
insensitive to the helicities of the pump and probe pulses. 
This behavior can be attributed to the orientation of the chiral Weyl nodes along the $k_{z}$ 
direction, which coincides with one of the field components of the driving pump and probe pulses.

The observed dependence on the polarization plane can be attributed to the alignment of the axis connecting the chiral Weyl nodes with the propagation direction of the pulses, i.e., the $z$-axis. 
In this configuration, the chirality of the Weyl nodes does not significantly influence the handedness of the circularly polarized driving fields. 
This behavior contrasts strongly with the case where the pulses are polarized in the $x-y$ plane, for which changing the helicity of either the pump or probe field leads to pronounced modifications in the harmonic and sideband spectra as shown in Fig.~\ref{fig2}. 
These results suggest that an orthogonal orientation between the node-separation axis and the driving field's polarization plane enhances the chiral response, enabling a robust coupling between Weyl node chirality and optical helicity.

\section{Conclusion}
In summary, we have investigated frequency mixing and helicity-dependent nonlinear optical response 
in an inversion-symmetric WSM driven by circularly polarized pump-probe pulses. 
The interaction of bicircular fields gives rise to pronounced sidebands originating from nonlinear mixing of pump and probe photons, enabling the generation of tunable high-harmonic spectra beyond conventional odd-order harmonics. 
We show that both the intensity and polarization characteristics of the generated sidebands depend sensitively on the relative helicities of the pump and probe pulses, as well as on the orientation of their polarization plane with respect to the Weyl nodes. 
This pronounced helicity dependence originates from the interplay between the chiral electronic structure of the Weyl nodes and the underlying rotational symmetry of the Weyl semimetal.
The appearance and suppression of specific sidebands can be understood through symmetry-allowed and symmetry-forbidden frequency mixing channels.
Furthermore, we established that while primary harmonics remain robust, the sideband intensities are acutely sensitive to femtosecond-scale temporal delays and electron-hole coherence times. 
This sensitivity positions helicity-dependent frequency mixing as a powerful spectroscopic tool for resolving ultrafast decoherence in topological systems.
Additionally, we show that the nonlinear response depends on the polarization plane of the driving fields relative to the axis connecting the Weyl nodes, highlighting the interplay between light helicity and chirality of the Weyl nodes.
The demonstrated helicity-dependent frequency mixing provides new opportunities for controlling nonlinear optical responses using tailored  light fields. 
Extending these ideas to multi-Weyl systems,  and Floquet-engineered quantum materials may enable new routes for ultrafast control of topological phases and lightwave-driven quantum technologies.
The ability to encode symmetry, chirality, and coherence into nonlinear optical signals suggests promising routes toward topological photonics, quantum information processing, and ultrafast optoelectronic functionalities based on emergent quantum materials.

\section{Acknowledgements}
We acknowledge  Leon Schlemmer from Paris for useful discussion. P.V. acknowledges the financial support from Anusandhan National Research Foundation (ANRF), India (NPDF Project no. PDF/2023/001102). 
 
The data that support the findings of this article are openly available ~\cite{thakur_2026}.
 

%

\end{document}